# Full Nonlinear Conversion of Broadband Frequency Combs generated by Four-Wave Mixing in Highly Nonlinear Fibers

**Arismar Cerqueira S. Jr, Jorge D. Marconi, Hugo L. Fragnito, and Flavio C. Cruz**

*Gleb Wataghin Physics Institute, University of Campinas - UNICAMP*
*13083-970, Campinas, SP, Brazil*
*flavio@ifi.unicamp.br*

**Abstract:** We generate a 200-nm optical frequency comb at 1.5 µm by injecting two single-frequency lasers into a highly nonlinear fiber, and explore its spectral phase properties and high power to demonstrate full nonlinear conversion into its second harmonic. Combs of optical frequencies with 100-nm (~ 48 THz) bandwidths, centered at 790-800 nm and spaced by 0.8-1 nm (380-470 GHz) were produced by second harmonic and multiple sum frequency generation simply by focusing into a 2-mm long nonlinear crystal. Conventional birefringence phase matching was used in normal incidence configuration without tilting the crystal.

## 1. Introduction

Broadband combs of highly coherent optical frequencies are generated today by mode-locked lasers. Originally employed to measure optical frequencies [1], they allowed the development of optical atomic clocks [2,3] and several other advances ranging from tests of possible variations of fundamental constants [4], generation of attosecond pulses [5], and phase-sensitive nonlinear optics [6]. Essential features in these combs are the stabilization of their two degrees of freedom: the frequency spacing (or pulse repetition rate) and the carrier-to-envelope offset frequency (or the pulse-to-pulse phase). Other recently demonstrated type of optical frequency comb, with bandwidths as large as 500 nm at 1550 nm, has been produced by nonlinear processes when a continuous-wave (cw) laser is injected into a high-Q silica microresonator [7]. The stabilization of these combs has been demonstrated [8], allowing applications in precision measurements and metrology.

Another type of broadband optical frequency comb can be generated by multiple four-wave-mixing (FWM) when two cw, single-frequency lasers are injected into a highly nonlinear optical fiber (HNLF). These combs have been studied with a main interest as a source of short pulses at high repetition rates for optical time-division multiplexing in optical telecommunication [9,10,11]. Recently, such combs with 300-nm bandwidth have been generated using very short lengths of HNLF [12]. Some of their distinct features are: 1) the absence of an optical resonator, which gives great flexibility in setting and changing the frequency spacing [13], 2) high-power, pulsed operation with peak powers of several Watts [12], 3) the possibility of obtaining flat spectral phase, leading to transform-limited pulses [10], and 4) a fixed pulse-to-pulse phase equal to $\pi$, arising from the original beatnote between the two input lasers [13]. Such "FWM-fiber-combs" benefit from the telecom technology, with advantages such as long-term and robust operation, compact sizes and moderate costs. Their potential use in precision applications still needs to be investigated, possibly requiring frequency (or phase) and power stabilization of the input lasers [14]. In this letter we report the generation of a 200-nm broadband comb at 1550 nm and demonstrate full nonlinear spectral conversion into its second harmonic by using birefringence phase-matching in a 2-mm thick BIBO crystal. The bandwidth of the second harmonic spectra of short pulse lasers with large spectral bandwidths is usually severely limited by the spectral phase of the fundamental pulses, requiring previous correction (e.g., pulse compression) by proper techniques [15]. Therefore our broadband conversion seems to be also a consequence of the small spectral phase variation of the comb. This is an important feature, which combined with an eventual increase in coherence and frequency stability provided by use of stabilized input lasers [14], may possibly lead to further applications such as high-bandwidth spread spectrum communication with increased security [16] and arbitrary waveform generation [17]. Our frequency converted, broadband light source at 800 nm can already be used for optical coherence tomography [18].

## 2. Broadband combs generated by four-wave mixing in highly nonlinear fibers

Figure 1 shows the experimental setup, similar to that used in ref. [12], to produce our FWM-fiber-comb. Two commercial, extended-cavity telecommunication semiconductor lasers provided cw, single-frequency output powers of 10 dBm each, and could be tuned through S, C and L-bands, from 1440 to 1620 nm. The two lasers are combined and amplified by two cascaded Erbium Doped Fiber Amplifiers (EDFAs, represented as a single one in Fig. 1). The first EDFA is a pre-amplifier with a typical average output power of 15 mW, whereas the second one is a booster with average output power up to 1 W. In order to obtain high peak powers, these pump signals passed through an amplitude modulator, which produced pulses typically 40-ns long with low duty cycle (7 μs separation). In our experiments we used peak powers of ~ 8 W for each laser. Since pulsed sources were used, the transient properties of the stimulated Brillouin scattering (SBS) should be considered [19]. Effects such as pulse shape variations or pulse narrowing [19] were not observed during the experiments (the pulse shape was continually monitored, Fig.1). We have also used a phase-modulator in order to evaluate the influence of SBS. The multiple FWM spectra, however, did not present any noticeable change when the pump linewidths were intentionally broadened by the phase modulator, indicating that SBS was not relevant. The pulses were then sent to two segments of low-dispersion HNLFs (from Sumitomo), and the generated spectrum was analyzed in an optical spectrum analyzer. The parameters for the first and second fiber segments were, respectively: lengths $L_1 = 15$ m and $L_2 = 5$ m; nonlinear coefficients $\gamma_1 = 15$ $W^{-1}$ $km^{-1}$ and $\gamma_2 = 10$ $W^{-1}$ $km^{-1}$; zero-dispersion wavelengths $\lambda_{01} = 1570$ nm and $\lambda_{02} = 1530$ nm; dispersion slopes $S_{01} = 0.015$ $ps/nm^2/km$ and $S_{02} = 0.02$ $ps/nm^2/km$.

Short pulses generated by FWM-fiber-combs have been particularly studied in a series of papers, by the group of G. Millot, in France. They reported the generation of well separated pulses with no pedestal, at rates extending from 20 GHz to up to 1 THz, and with durations from 280 fs to a few ps [11,13,20]. They also performed extensive characterization of the pulses, which included spectral phase and pulse duration measurements using frequency-resolved optical gating (FROG) and auto-correlation. The generated pulses can be interpreted as a compression of the original beatnote produced by the two input lasers. The phase across the pulse can be constant, leading to transform-limited pulses and constant spectral phase, provided that the HNLF length and optical power have optimum values. For short lengths of low-dispersion fibers, the phase variation across the spectrum can be quite small, which will be important for second harmonic generation.

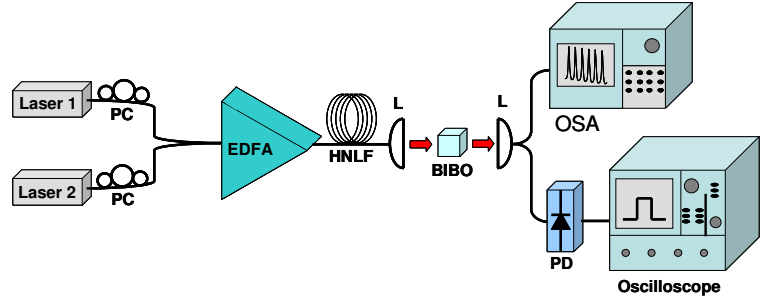

Fig. 1. Experimental setup used to generate an OFC by multiple FWM in a highly nonlinear fiber. Lasers 1 and 2 are standard telecom extended-cavity semiconductor lasers. PC: polarization controller, EDFA: Erbium doped fiber amplifier, HNLF: highly nonlinear fiber; L:lens; BIBO: nonlinear BIBO crystal; PD: photodetector; OSA: optical spectrum analyzer.

## 3. Broadband second harmonic generation

Figure 2 shows the full nonlinear conversion of the comb by second harmonic generation. Second harmonic or sum frequency generation between two single-frequency lasers require only matching of phase velocities, which can be achieved by birefringence or temperature tuning. For broadband light sources, such as femtosecond lasers or a FWM-fiber-comb, broadband sum-frequency generation requires, in addition, also matching of the group velocities and the group velocity dispersion (GVD) [21]. For short pulses (<100 fs), with large bandwidths, matching of high-order dispersion terms, such as cubic and fourth order dispersion, is also required. For this reason thin crystals are usually employed for broadband input sources (or ultrashort pulses), minimizing group velocity mismatch but also leading to smaller efficiencies. Even for thin crystals, the bandwidth of second harmonic generation is limited by the spectral phase of the fundamental laser. Since each generated frequency at 2f results from the simultaneous contribution of all frequency pairs symmetrically displaced from f, if the phase varies along the spectrum, each of those pairs will produce its sum-frequency with a corresponding different phase, leading to destructive interference and low output power. If however the spectral phase is constant, corresponding to TL pulses, thin crystals can generate the broadest second harmonic spectrum. Ref. [15] demonstrated a 110-nm bandwidth second harmonic spectrum generated in a 200 μm KDP crystal by a broadband mode-locked laser, whose spectral phase had to be corrected to produce TL pulses. If the spectral phase is not constant, or has significant variations, the SH bandwidth will also be limited.

As shown in Figure 1, the setup for second harmonic and sum frequency generation consists simply of focusing the output of the HNLF into a nonlinear crystal. We employed 2-mm and 100-μm long crystals (AR-coated, cut for critical phase matching at room temperature, with $\theta = 10.9^0$, $\phi = 0^0$), whose SH spectra are plotted in figure 2, together with the corresponding fundamental spectra. For SHG and SFG of 1.55 μm, BBO is a good choice since it has an extremely large phase matching bandwidth [22] (600 nm is calculated for a 10 mm long crystal [23]). We however chose BIBO, another crystal for the same family, whose phase-matching bandwidth at 1.55 μm is 250 nm for our 2-mm long crystal [23]. Compared to BBO, it has higher (9%) nonlinear coefficient, larger (48%) acceptance angle and less (60%) walk-off than BBO. The second harmonic power was estimated from the incident powers and the crystal efficiency. For fundamental peak powers of 8 W at $\omega_1 = 2\pi c/1562$ nm and $f_2 = 2\pi c/1565$ nm (Fig.1), the maximum second harmonic peak power at 782 nm is estimated from the nonlinear coefficient ($\eta = 3.7 \times 10^{-5}$ W$^{-1}$ for the 2-mm BIBO crystal), under optimum focusing conditions [24], as $P_{(\omega1+\omega2)} = \eta P_{\omega1} P_{\omega2} = 2.4$ mW. The estimated average power is 14 µW, for 40 ns pulses separated by 7 µs. In fig. 2a, 61 fundamental frequencies (including the pump lasers) generate 122 comb lines centered at 790 nm and separated by 0.8 nm (380 GHz) in a 100-nm bandwidth. In fig. 2b (for the 100 µm crystal), a SH bandwidth of 120 nm is achieved, with frequency spacing of ~ 1 nm (480 GHz). The frequency spacing in the fundamental spectrum is preserved in the second harmonic because each fundamental frequency f produces its second harmonic at 2f, but each frequency pair symmetrically located around f produces a sum frequency in between. In addition, each SH frequency also has contributions from multiple SFG between many line pairs. Higher nonlinear conversion efficiency can possibly be obtained by using longer crystals, while still keeping large phase-matching bandwidths. For comparison, the efficiency of second harmonic generation (SHG) in combs produced in silica microresonators might be largely limited by the relatively low fundamental power.

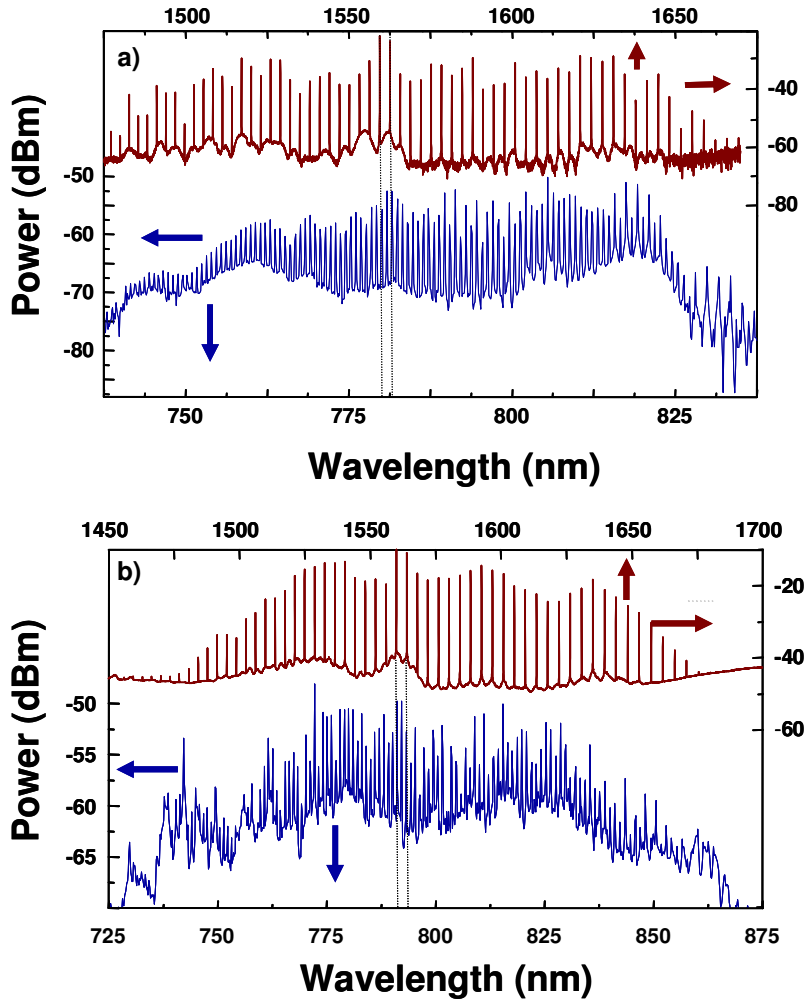

Fig. 2. Spectra of FWM-fiber-combs at 1550 nm, and their second harmonic spectra. Figures a and b correspond to the 2-mm and 0.1-mm long nonlinear BIBO crystal. The collection efficiency for the second harmonic was far from optimized. Dotted vertical lines indicate the input lasers.

An extension of the spectral coverage of the fundamental comb can be expected either from optimizing the fiber-laser configuration, by cascading more stages of nonlinear mixing in different fibers, or by phase-locking separate combs with different spectra. The spectral conversion of a comb produced with mature telecom technology at 1.5 μm into other regions where technology is either expensive or lacking is of particular importance. Possible future developments include nonlinear conversion into the infrared region (2-6 microns), where strong absorption bands of several gases are present. This, which can possibly be achieved by second-order processes such as difference frequency generation or by parametric down-conversion, will allow new opportunities for molecular fingerprinting and real-time, high-sensitivity spectroscopy of multiple gases in a mixture. Our broadband light source at 790 nm, based on telecom technology, can also be a convenient alternative for optical coherence tomography.

## 4. Conclusion

We demonstrated full nonlinear conversion of a 200-nm optical comb at 1550 nm into its second harmonic. Combs of optical frequencies with 100-nm bandwidths, centered at 800 nm and spaced by 0.8-1.0 nm have been produced. The broadband conversion is a consequence of the large phase-matching bandwidth of the BIBO crystal and the small phase variation across the fundamental comb spectrum. The comb frequency spacing is preserved upon nonlinear conversion due to sum-frequency mixing, which creates intermediate frequencies in the comb spectrum. Prospects for nonlinear conversion into the infrared (2-6 microns) may open new opportunities for molecular fingerprinting and broadband high resolution spectroscopy.

**Acknowledgments**: Financial support from FAPESP, CEPOF, and CNPq is gratefully acknowledged. FCC acknowledges stimulating discussions and suggestions from Prof. Marcos Dantus, from Michigan State University. Corresponding author: F. C. Cruz (flavio@ifi.unicamp.br )